\documentclass[runningheads]{llncs}
\usepackage{amsmath,amssymb,amsfonts}
\usepackage{algorithmic}
\usepackage{graphicx}
\usepackage{textcomp}
\usepackage{boldline}
\usepackage{multicol,multirow}
\usepackage{adjustbox}
\begin{document}
\title{Only-Train-Once MR Fingerprinting for Magnetization Transfer Contrast Quantification}
\titlerunning{Only-Train-Once MR Fingerprinting (OTOM)}
%
\author{Beomgu Kang\inst{1} \and
Hye-Young Heo\inst{2} \and
HyunWook Park\inst{1}}
\authorrunning{B. Kang et al.}
%
\institute{School of Electrical Engineering, Korea Advanced Institute of Science and Technology, Daejeon, Republic of Korea\\\and
Divison of MR Research, Department of Radiology, Johns Hopkins University, Baltimore, Maryland, USA\\
}
\maketitle              
\begin{abstract}
Magnetization transfer contrast magnetic resonance fingerprinting (MTC-MRF) is a novel quantitative imaging technique that simultaneously measures several tissue parameters of semisolid macromolecule and free bulk water. In this study, we propose an Only-Train-Once MR fingerprinting (OTOM) framework that estimates the free bulk water and MTC tissue parameters from MR fingerprints regardless of MRF schedule, thereby avoiding time-consuming process such as generation of training dataset and network training according to each MRF schedule. A recurrent neural network is designed to cope with two types of variants of MRF schedules: 1) various lengths and 2) various patterns. Experiments on digital phantoms and \textit{in vivo} data demonstrate that our approach can achieve accurate quantification for the water and MTC parameters with multiple MRF schedules. Moreover, the proposed method is in excellent agreement with the conventional deep learning and fitting methods. The flexible OTOM framework could be an efficient tissue quantification tool for various MRF protocols.

\keywords{Deep learning \and MR fingerprinting  \and Magnetization Transfer Contrast \and Recurrent neural network \and Transfer Learning.}
\end{abstract}
\section{Introduction}
Magnetization transfer contrast (MTC) imaging is a MRI technique that provides information about semisolid macromolecular protons, based on the transfer of their magnetization to surrounding free bulk water \cite{RN20,RN22,RN21}. Since the macromolecules are not directly detectable in MRI, due to the extremely short T\textsubscript{2} ($<$100 $\mu$T), the continuous saturation of solute protons with radiofrequency (RF) irradiation and the subsequent transfer of proton allow us to assess the macromolecular protons indirectly from the reduced water signal. In traditional MTC experiments, magnetization transfer ratio (MTR), the ratio between two images acquired with and without RF saturation, has been widely used in the clinic \cite{RN19,RN24}. Nonetheless, the qualitative nature of MTR metric, including its high dependency on scan parameters and tissue relaxation effects, limits its clinical usefulness.

To address this issue, magnetic resonance fingerprinting (MRF) technique was introduced in MTC imaging as a promising and time efficient technique that simultaneously quantified multiple tissue parameters \cite{RN16,RN5,RN1}. In MRF acquisition, time-varying signal evolution is achieved by intentionally varying imaging parameters for each scan. The acquired evolution is the distinguishing characteristics for a set of tissue parameters, so called a fingerprint. Then, the acquired fingerprints of a voxel are mapped with a pre-calculated dictionary. However, dictionary matching based techniques suffer from an exhaustive search and the dictionary discretization issue. Recent studies in deep learning based MRF techniques have showed their powerful ability to map a MR fingerprint space to a tissue parameter space, greatly accelerating the reconstruction and improving the quantification accuracy \cite{RN2,RN3,RN4,RN7}. 

\begin{figure}[!b]
\includegraphics[width=\textwidth]{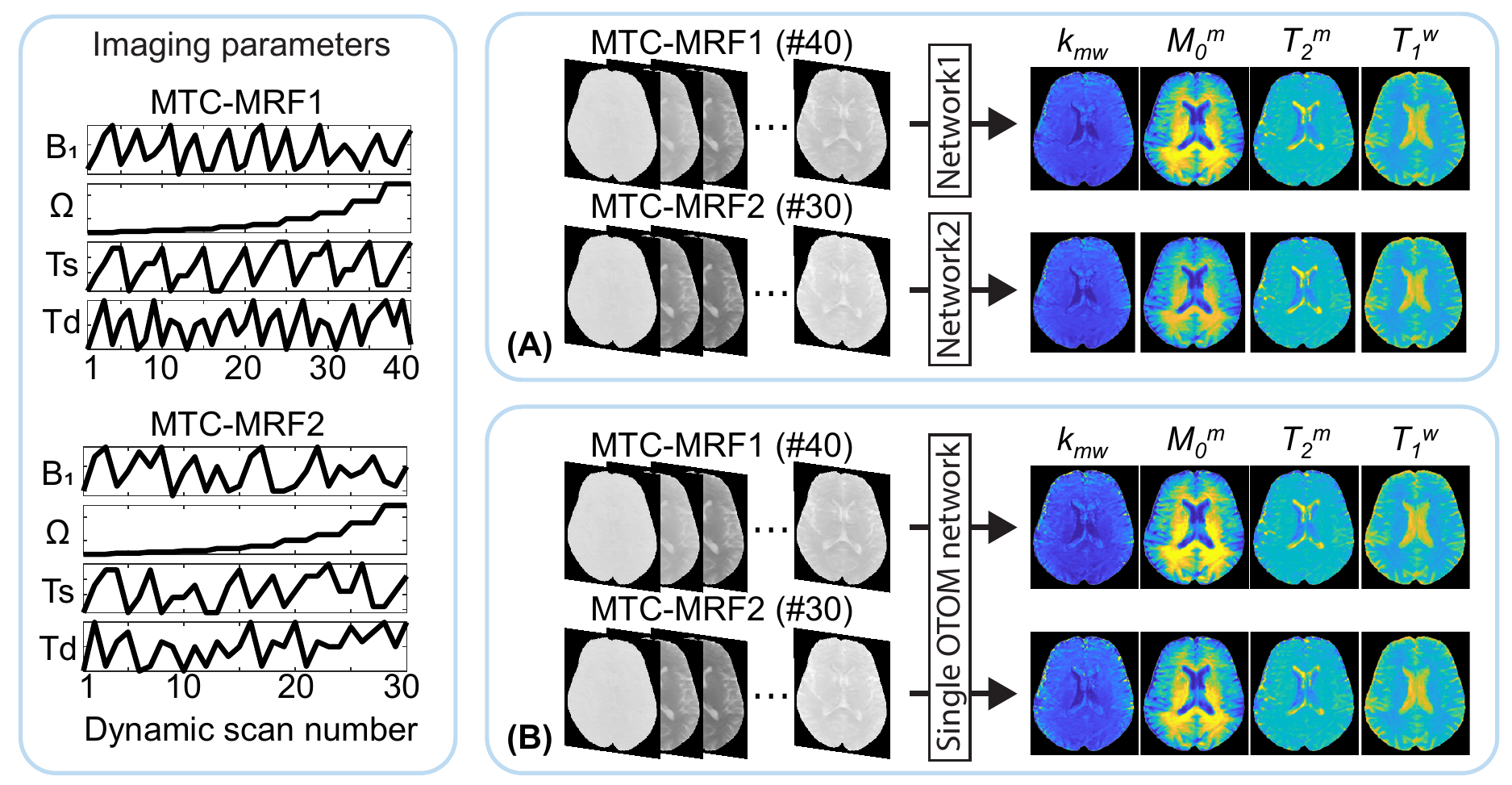}
\caption{(A) The original deep learning MRF framework. (B) The proposed framework.} \label{fig1}
\end{figure}

However, the deep neural networks were constrained to a single MRF schedule corresponding to the training dataset. If the MRF schedule is changed, the deep neural network has to be trained with new training dataset that was generated with the new MRF schedule. This process is very time consuming and inefficient. Therefore, the utility of MRF techniques would benefit greatly from the development of streamlined deep learning frameworks or even only-train-once methods for various MRF sequences. Recently, to streamline the deep learning process, dataset generation was accelerated using a parallel execution of large-scale MR signal simulations with exact MR physics models on graphics processing units (GPUs) \cite{RN8,RN9}. To further accelerate the simulations, deep neural networks have been proposed as surrogate physics models for computing MR signals \cite{RN10,RN11,RN12}. Although the simulation-accelerated approach can largely reduce the time complexity of MR signal models, re-training of the deep neural network remains to be a challenge. 

In this study, we propose an Only-Train-Once MR fingerprinting (OTOM) framework that maps a MR fingerprint space and a scan parameter space into a tissue parameter space. The proposed method can be applied to any MRF schedules unlike the previous deep learning based studies dedicated to only a single fixed MRF schedule.

\section{Methods}
\subsection{Signal model: Transient-state MTC-MRF}
A two pool exchange model, including the free bulk water pool (w) and the semisolid macromolecule pool (m), is used to simulate the MTC-MRF signal in the presence of proton exchange and RF irradiation. The magnetization of each pool can be described with the modified Bloch-McConnell equations. By solving the coupled differential equation, the transient-state MTC-MRF signal evolution ($S_{MTC-MRF}$) can be described using the following signal equation \cite{RN13,RN14}:
\begin{multline}
S_{MTC-MRF}(T_1^w,T_1^m,T_2^w,T_2^m,k_{mw},M_0^m;B_1,\Omega,Ts,Td) \\
=\left[M_0^w\left(1-e^{-Td/T_1^w}\right)-M_{ss}^w\right]e^{-\lambda Ts}+M_{ss}^w
\label{eq1}
\end{multline}
where $T_1^i$ and $T_2^i$ are the longitudinal and transverse relaxation times of a pool i, respectively; $k_{ij}$ is the proton exchange rate from a pool i to a pool j; $M_0^i$ is the equilibrium magnetization of a pool i; $M_{ss}^w$ is the steady-state longitudinal magnetization of the free bulk water; $B_1$ is the RF saturation power; $\Omega$ is the frequency offset of the RF saturation; Ts is the saturation time; Td is the relaxation delay time; and $\lambda$ represents the longitudinal relaxation rates of the water pool under the saturation of the macromolecule pool.

Deep learning was incorporated with the analytical solution of the MTC-MRF signal model to understand complex relation among the fingerprint space ($S_{MTC-MRF}$), the scan parameter space ($B_1$, $\Omega$, Ts, and Td) and the tissue parameter space ($k_{mw},M_0^m,T_2^m$ and $T_1^w$). 

\begin{figure}[!t]
\centerline{\includegraphics[width=\columnwidth]{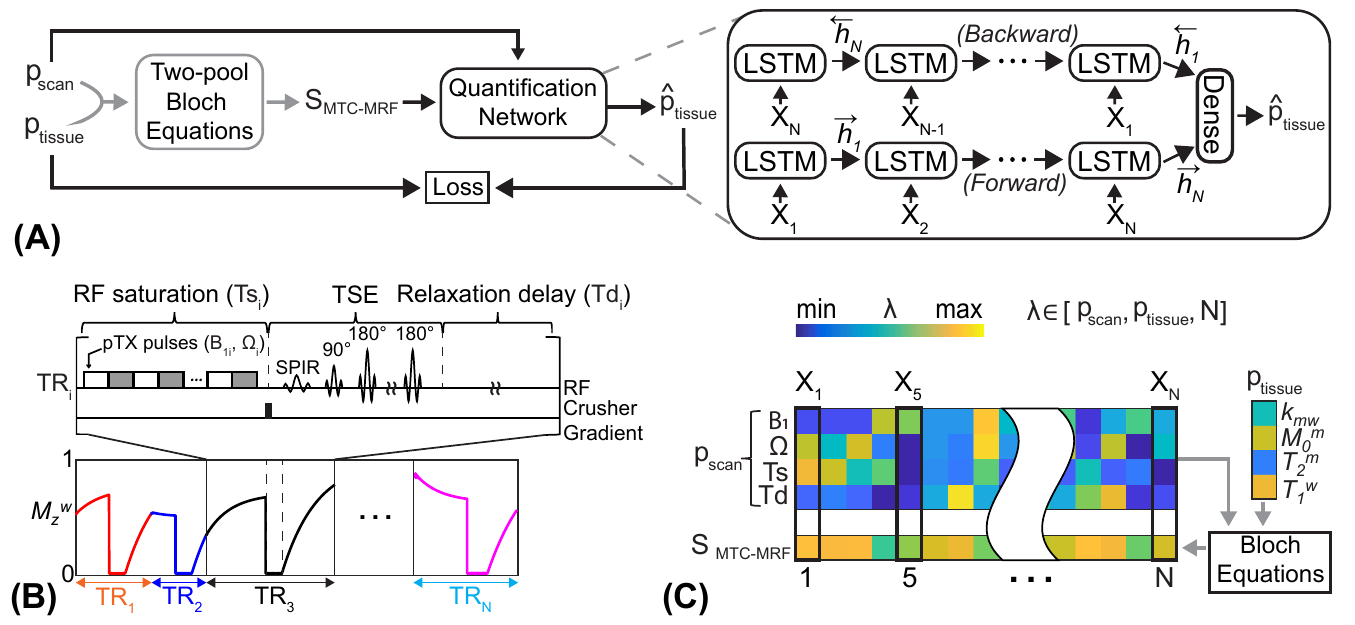}}
\caption{An overview of the OTOM framework. (A) A recurrent neural network for the water and MTC quantification. (B) The evolution of the longitudinal magnetization of water according to the turbo spin-echo (TSE)-based MTC-MRF schedule. (C) The length (N), scan parameters (p\textsubscript{scan}), and tissue parameters (p\textsubscript{tissue}) are randomly sampled within the pre-defined range and used to generate MRF signal through Bloch equations.}
\label{fig2}    
\end{figure}

\subsection{Proposed Model}
\subsubsection{Recurrent neural network (RNN)}
The RNN architecture was designed to quantify the free bulk water and semisolid MTC parameters from given MTC fingerprints and the corresponding scan parameters. In specific, a bi-LSTM (bidirectional long short-term memory) \cite{RN15} architecture was used with a single fully connected layer (Fig.\ref{fig2}A). The bi-LSTM processes an input sequence in two ways: moving forward from the start to the end of the sequence, and moving backward from the end to the start of the sequence. LSTM extracts features from each time point of the input sequence and accumulates the features in the form of a hidden state. Forward and backward hidden states at the end are concatenated and fed to the dense layer to estimate the four tissue parameters ($p_{tissue}$) of $k_{mw},M_0^m,T_2^m,$ and $T_1^w$, as follows:
\begin{equation}
\overrightarrow{h}_i=R\left(X_i,\overrightarrow{h}_{i-1};\theta\right)
\label{eq2}\end{equation}
\begin{equation}
\overleftarrow{h}_i=R\left(X_i,\overleftarrow{h}_{i+1};\theta\right)
\label{eq3}\end{equation}
\begin{equation}
\widehat{p}_{tissue}=f\left(\overrightarrow{h}_N,\overleftarrow{h}_1\right)
\label{eq4}\end{equation}
\begin{equation}
X_i=\left[S_{MTC-MRF},B_1,\Omega,Ts,Td\right]_i,\;i=1,\cdots,N
\label{eq5}\end{equation}
\noindent
where $X_i$ is the input vector of $i^{th}$ time point; $\overrightarrow{h}_i$ and $\overleftarrow{h}_i$ are the forward and backward hidden states, respectively, at $i^{th}$ time point; $\theta$ represents the parameters of the LSTM; $R$ denotes the LSTM network; $f$ is the dense layer followed by rectified linear units (ReLU); and $N$ is the number of dynamic scans of the MRF sequence. The LSTM consists of three layers with 512 hidden units. The RNN architecture was trained by minimizing the absolute difference between the label parameter $p_{tissue}$ and the estimated parameter $\widehat{p}_{tissue}$, as follows: 
\begin{equation}
L_\theta=\left|p_{tissue}-\widehat{p}_{tissue}\right|
\label{eq6}\end{equation}
The RNN network was implemented using Pytorch 1.8.1 on an NVIDIA TITAN RTX GPU (Santa Clara, CA). The adaptive moment estimation (ADAM) optimizer \cite{RN23} was used with the initial learning rate of 10\textsuperscript{-3} and a batch size of 256. The learning rate was scheduled to be decreased by a factor of 0.1 for every 5 epochs. The training data was randomly divided into two parts: 90\% for training and 10\% for validation. The validation loss was used for model selection and early stopping.

\subsubsection{Dataset generation}
The analytical solution of Bloch-McConnell equations (Eq. \eqref{eq1}) was used to generate MTC-MRF signal ($S_{MTC-MRF}$) for given tissue parameters and scan parameters. To train the RNN model that successfully processes random MRF schedules, two variants of MRF schedules with respect to the length and the pattern were considered (Fig. \ref{fig2}C). For scan parameters $p_{scan}$=[$B_1$, $\Omega$, Ts, Td], 
\\1) The length of MRF ($N$) was randomly selected within the range of 10 to 40. 
\\2) The MRF schedule s = [$p_{scan,1}$, $p_{scan,2}$,…, $p_{scan,N}$] was randomly sampled, where every scan parameter was constrained to the pre-defined range: 8–50 ppm for frequency offsets ($\Omega$); 0.5–2.0 $\mu$T for RF saturation power ($B_1$); 0.4–2.0 s for RF saturation time (Ts); and 3.5–5.0 s for relaxation delay time (Td).
\newline\noindent
For tissue parameters $p_{scan}$ = [$k_{mw},M_0^m,T_2^m,T_1^w$], each parameter was randomly sampled. The ranges of tissue parameters were: 5–100 Hz for $k_{mw}$; 2–17\% for $M_0^m$; 1–100 $\mu$s for $T_2^m$; and 0.2–3.0 s for $T_1^w$. To generate the training dataset, 80 million combinations of scan and tissue parameters were chosen as explained. Finally, white Gaussian noise (SNR=46dB) was added to the simulated MTC-MRF signals. The Gaussian noise level was determined from the estimated SNR of the acquired \textit{in vivo} images.

\subsubsection{Transfer learning (TL)}
To gauge how the OTOM is well optimized for each schedule, TL was used. The OTOM is a baseline container model that can processes any given MRF schedule. Therefore, TL to a specific MRF schedule was available to further optimize the network. For the training dataset, ten million sets of tissue parameters were randomly sampled with the target MRF sequence. The OTOM-T (OTOM-Transfer) network was trained for 3 epochs with the adaptive moment estimation (ADAM) optimizer. The learning rate was scheduled to be decreased by a factor of 0.1 for every 2 epochs and the initial rate was 10\textsuperscript{-4} with a batch size of 256. 

\section{Experiments}
\subsection{Digital phantom study: Bloch simulation}
\begin{figure}[!t]
\centerline{\includegraphics[width=\columnwidth]{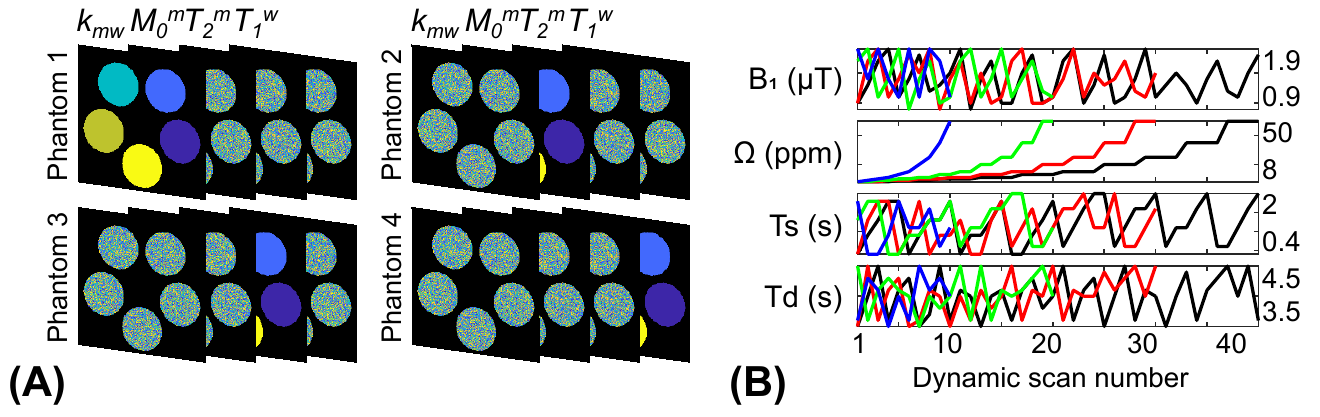}}
\caption{Bloch-McConnell equation-based digital phantom study. (A) Ground truths for four digital phantoms (B) Four pseudo-random (PR) schedules with different numbers of dynamic scans (N = 10, 20, 30, and 40) were used to simulate MTC-MRF images.}
\label{fig3}    
\end{figure}

The performance of the OTOM framework was validated with the Bloch-McConnell equation-based simulation (Fig. \ref{fig3}). Four digital phantoms were constructed to have five uniformly sampled constant values for a single parameter while the other parameters were randomly sampled within the pre-defined ranges. Five uniformly sampled values for each phantom were: 5, 25, 50, 75, and 100 Hz for $k_{mw}$, 2, 6, 10, 14, and 17 \% for $M_0^m$, 1, 25, 50, 75, and 100 $\mu$s for $T_2^m $, and 0.2, 0.9, 1.6, 2.3, and 3.0 s for $T_1^w$. The quantification accuracy was evaluated using pseudo-random (PR) schedules with various numbers of dynamic scans (N = 10, 20, 30, and 40) by calculating mean absolute errors (MAE). Reconstruction results of the OTOM method were compared with those of the Bloch fitting \cite{RN16} and FCNN-based deep learning approach \cite{RN3}. For the FCNN method, four different FCNN networks were respectively trained for the corresponding four PR schedules. The PR schedules were generated by minimizing the information redundancy between multiple dynamic scans \cite{RN6}.

\begin{table}[!t]
\caption{Quantitative evaluation of mean absolute error (MAE) from OTOM, FCNN and Bloch fitting methods for Bloch-McConnell equation-based digital phantoms.}
\centering
\begin{tabular}{p{2.5cm} p{2.5cm} p{1.6cm} p{1.6cm} p{1.6cm} p{1.6cm}}
\hlineB{3}
Schedule & Method & \multicolumn{4}{c}{MAE}\\ \cline{3-6}
with length &  & $k_{mw} (Hz)$ & $M_0^m (\%)$ & $T_2^m (\mu s)$  & $T_1^w (ms)$ \\ \hlineB{1.5}
\multirow{3}{*}{PR\#40} &OTOM          & 10.29 & 0.581 & 1.501 & 21.70 \\ \cline{2-6}
                        &FCNN          & \textbf{10.09} & \textbf{0.547} & \textbf{1.385} & \textbf{20.75} \\ \cline{2-6}
                        &Bloch fitting & 12.07 & 0.735 & 2.437 & 24.06 \\\hlineB{1.5}
                     
\multirow{3}{*}{PR\#30} &OTOM          & 11.19 & 0.658 & 1.708 & 23.23 \\ \cline{2-6}
                        &FCNN          & \textbf{11.18} & \textbf{0.644} & \textbf{1.638} & \textbf{22.89} \\ \cline{2-6}
                        &Bloch fitting & 13.10 & 0.842 & 2.728 & 26.74 \\\hlineB{1.5}     
                     
\multirow{3}{*}{PR\#20} &OTOM          & 11.84 & 0.783 & 2.044 & 31.81 \\ \cline{2-6}
                        &FCNN          & \textbf{11.78} & \textbf{0.778} &\textbf{1.971} & \textbf{31.50} \\ \cline{2-6}
                        &Bloch fitting & 13.70 & 1.037 & 3.725 & 43.53 \\\hlineB{1.5}
                     
\multirow{3}{*}{PR\#10} &OTOM          & \textbf{13.91} & 0.980 & 2.713 & 42.37 \\ \cline{2-6}
                        &FCNN          & 13.94 & \textbf{0.973} & \textbf{2.689} & \textbf{42.10} \\ \cline{2-6}
                        &Bloch fitting & 16.10 & 1.325 & 4.896 & 61.85 \\                   
\hlineB{3}
\end{tabular}
\label{tab1}
\end{table}

\subsection{In vivo Experiments}
Six healthy volunteers (M/F: 2/4; age: 36.2 ± 3.7 years) were scanned on a Philips Achieva 3T MRI system with the approval of the institutional review board at Johns Hopkins university, and written informed consent was obtained prior to the MRI experiments. The 3D MTC-MRF images were acquired from a fat-suppressed (spectral pre-saturation with inversion recovery) multi-shot TSE pulse sequence. All image scans were obtained with 4× compressed sensing (CS) accelerations in the two phase-encoding directions (ky-kz) \cite{RN17}. The imaging parameters were: TE = 6 ms; FOV = 212 × 186 × 60 mm$^3$, spatial resolution = 1.8 × 1.8 × 4 mm$^3$, slice-selective 120$^\circ$ refocusing pulses, turbo factor = 104, and slice oversampling factor = 1.4. The two-channel time-interleaved parallel RF transmission (pTX) technique was used to achieve continuous RF saturation of 100\% duty-cycle by distributing the saturation burden into two amplifiers \cite{RN18}. Forty dynamic scans were acquired with various RF saturation powers ($B_1$), frequency offsets ($\Omega$), RF saturation times (Ts), and relaxation delay times (Td) according to the PR schedule. To normalize the MTC-MRF images, an additional unsaturated image ($S_0$) was acquired. 

\begin{figure}[!t]
\centerline{\includegraphics[width=\columnwidth]{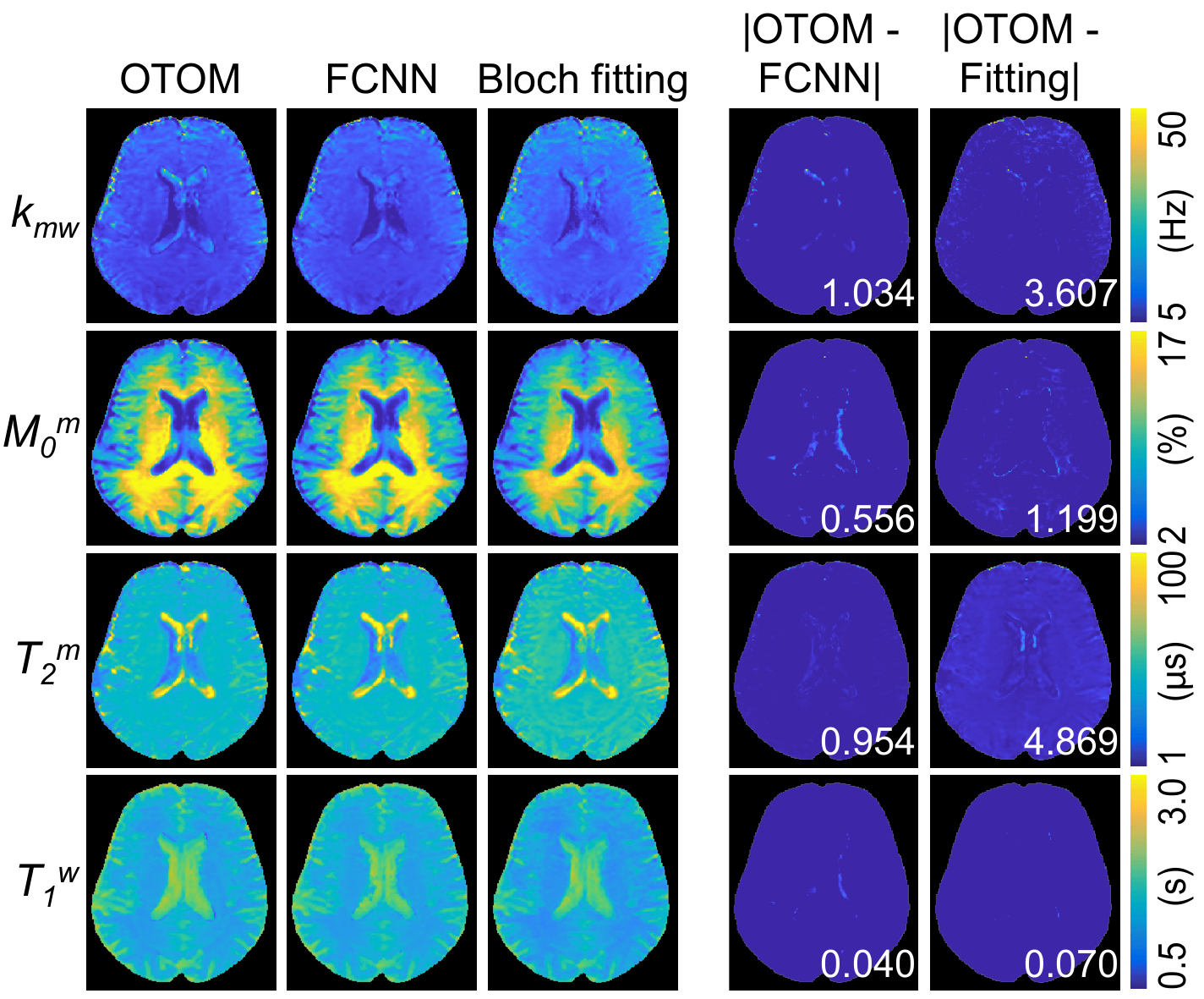}}
\caption{Representative MTC parameters and water T\textsubscript{1} map estimated using the OTOM, FCNN, and Bloch fitting methods. Difference images of the estimated maps with OTOM versus FCNN and fitting methods. The mean difference value of each map is shown in the insert (white).}
\label{fig4}    
\end{figure}

\subsection{Results and Discussion}
As shown in Table \ref{tab1}, the OTOM and FCNN methods showed considerably similar quantification error. The OTOM framework successfully decoded fingerprints into tissue parameters irrespective of their MRF schedules. Moreover, both OTOM and FCNN methods showed a high degree of the reconstruction accuracy for every tissue parameter compared to the Bloch fitting approach. Moreover, both deep learning methods take advantage of low computational complexity for tissue parameter quantification due to their simple feed-forward deployments in test phase (110 sec for OTOM and 6 sec for FCNN with an image matrix of 256 x 256 x 9 x 40). In comparison, the Bloch fitting method suffers from the high computational cost for the tissue parameter mapping (one hour with an image matrix of 256 x 256 x 9 x 40). However, the FCNN approach is constrained to a specific MRF schedule and thus a time-consuming re-training process of neural networks is required for other MRF schedules, thereby limiting its clinical utility. On the contrary, the proposed OTOM can immediately cope with various MRF sequences.

Figure \ref{fig4} shows the excellent agreement between the estimated \textit{in vivo} maps from the OTOM method and those from the FCNN method (p $<$ 0.05 for all parameters; correlation coefficients of 0.89, 0.98, 0.99, and 0.96, respectively for $k_{mw},M_0^m,T_2^m$ and $T_1^w$). The estimated parameter maps from the fitting method were also in excellent agreement with those from the OTOM method (0.81, 0.99, 0.94, and 0.99, respectively for $k_{mw},M_0^m,T_2^m$ and $T_1^w$). 

\begin{table}[!t]
\caption{Quantitative evaluation of OTOM-T (OTOM-Transfer) and FCNN methods with PR \#40 schedule on Bloch-McConnell equation-based digital phantoms.}
\centering
\begin{tabular}{p{2.8cm} p{1.6cm} p{1.6cm} p{1.6cm} p{1.6cm} p{2.5cm} }
\hlineB{3}
Method     &  \multicolumn{4}{c}{MAE} & Data preparation\\ \cline{2-5}
 &  $k_{mw} (Hz)$ & $M_0^m (\%)$ & $T_2^m (\mu s)$  & $T_1^w (ms)$ & and training \\ \hlineB{1.5}

OTOM (Baseline)& 10.29 & 0.581 & 1.501 & 21.70 & -\\ \hline
OTOM-T & \textbf{10.06} & \textbf{0.545} & \textbf{1.343} & \textbf{19.79} & 170 (min)\\ \hline
FCNN & 10.09 & 0.547 & 1.385 & 20.75 & 370 (min)\\ \hline
\hlineB{3}
\end{tabular}
\label{tab2}
\end{table}

In addition, the OTOM-T network, achieved by transfer learning of the trained OTOM network to a certain MRF schedule, showed higher accuracy than the OTOM and FCNN methods (Table \ref{tab2}). Note that the same dataset was used to train the OTOM-T and FCNN. However, the performance gain from the use of transfer learning is small: the normalized root mean square errors (nRMSE) difference between OTOM and OTOM-T were 0.26, 0.42, 0.21, and 0.02 \% for $k_{mw},M_0^m,T_2^m$ and $T_1^w$. This refers that the OTOM network is well optimized for various schedules. 

\section{Conclusion}
We proposed an Only-Train-Once MR fingerprinting (OTOM) framework to estimate the free bulk water and MTC parameters from MR fingerprints regardless of the MRF sequence. Unlike the previous deep learning studies, the proposed method was trained with numerous patterns and lengths of MRF sequence, allowing us to plug any MRF sequence rather than a fixed MRF sequence. The flexible OTOM framework could be an efficient tissue quantification tool for various MRF protocols.

\subsubsection{Acknowledgement}
This work was supported by the Korea Medical Device Development Fund grant funded by the Korea government (the Ministry of Science and ICT, the Ministry of Trade, Industry and Energy, the Ministry of Health and Welfare, the Ministry of Food and Drug Safety) (Project Number: 1711138003, KMDF-RnD KMDF$\_$PR$\_$20200901$\_$0041-2021-02), and by grants from the National Institutes of Health (R01EB029974 and R01NS112242).

%
%
%
\bibliographystyle{splncs04}
\bibliography{ms}

\end{document}